\documentclass[
  journal=largetwo,
  manuscript=article-type,
  year=2020,
  volume=37,
]{cup-journal}

\usepackage{amsmath}
\usepackage{longtable}
\usepackage{amssymb}
\usepackage{hyperref}
\usepackage[nopatch]{microtype}
\usepackage{booktabs}
\usepackage{lscape}
\usepackage{longtable}
\usepackage{subfig}
\usepackage{supertabular,booktabs}
\usepackage{gensymb}
\usepackage{float}
\usepackage{lscape} 
\usepackage{aas-macros}
\title{Exploring LSST’s capabilities for early detection of outbursts in low-mass X-ray binaries}

\author{Susmita Sett}
\affiliation{International Centre for Radio Astronomy Research, Curtin University, Bentley, WA 6102, Australia}

\author{Arash Bahramian}
\affiliation{International Centre for Radio Astronomy Research, Curtin University, Bentley, WA 6102, Australia}

\author{Kristen Dage}
\affiliation{International Centre for Radio Astronomy Research, Curtin University, Bentley, WA 6102, Australia}

\author{David Russell}
\affiliation{New York University Abu Dhabi, Saadiyat Island, P.O. Box 129188, Abu Dhabi, United Arab Emirates}

\author{William I. Clarkson}
\affiliation{University of Michigan-Dearborn, 4901 Evergreen Rd, Dearborn, MI 48128, USA}


\begin{document}

\begin{abstract}

Following long periods of quiescence, low-mass X-ray binaries can exhibit intense X-ray outbursts triggered by instabilities within the accretion disk. These outbursts can sometimes be detected in optical wavelengths before being detected in X-ray, acting as an early onset warning and enabling a deep study of accretion disk properties informed by the lag between optical and X-ray rise. We explore the potential of Vera C. Rubin Observatory’s Legacy Survey of Space and Time (LSST) to detect these outbursts early through optical observations. We evaluate the capabilities of LSST based on currently planned survey cadence, filter-specific depth, and other observational factors that affect early detection. We develop and apply an extended metric to assess outburst detectability and recovery fraction. We find that despite inherent potential for early detection of XRB outbursts, the currently planned survey strategy makes it challenging to detect early onset of XRBs. Lastly, we demonstrate how this estimate can be used to infer the wider LMXB population in the Galaxy as the LSST progresses.
\end{abstract}

\section{Introduction} 
\label{sec:intro}

Low mass X-ray binaries (LMXBs) consist of a neutron star or black hole accreting matter from a low-mass ($\lesssim \rm  1M_\odot  $) donor star. LMXBs in our Galaxy generally remain in a state of quiescence (faint and at low accretion rates with X-ray luminosities $\sim 10^{29} - 10^{33} \rm erg~s^{-1}$) for years to decades. However, they can be easily detected in short (days to weeks) periods of outbursts, when the X-ray emission is much brighter \citep[up to $\sim 10^{39} \rm erg~s^{-1}$][and references therein]{ref:lmxbreview, ref:heinke, ref:tetarenko}. LMXB outbursts typically show a fast rise followed by an exponential decay \citep[e.g.,][]{ref:chen}. The initial rise of several days to weeks corresponds to complete ionization of the accretion disk and increase in mass transfer onto the compact object, followed by a long decay as energy is radiated cooling the ionized disk \citep[and references therein]{ref:lmxbreview}. 


Optical detection and monitoring offers an alternative pathway to the discovery and characterization of LMXBs. The initial stages of the outburst can be first observed at optical wavelengths several days {\it before} the X-ray outburst \citep{ref:russel}. As a recent example, the X-ray binary MAXI J1820+070 was discovered to be an X-ray binary due to its X-ray outburst \citep{ref:kawamuro}, however it was recorded as an optical transient, named, ASASSN-18ey a few days earlier \citep{ref:tucker}. 

Early pre-outburst detection via optical monitoring holds several advantages over detection via the X-ray outburst alone. Since the optical outburst rise precedes the X-ray brightening, this can enable earlier X-ray follow-up during the initial rise and study the mechanisms that trigger the outburst, which are not understood due to a relative scarcity of observations during the earliest phases of outburst onset. Moreover, the early detection of the outburst across the spectrum (X-ray and optical) enables us to constrain characteristics of the accretion disk. Furthermore, the detection of these systems in general can be useful in triggering more observations in the optical band and help in better characterisation of the light curve. With a larger and more thoroughly studied sample of LMXBs, we can gain deeper insights into the evolution of the outburst. These extended trends are particularly significant because they help differentiate the optical signals of these systems from those of other systems that also show ellipsoidal modulation like cataclysmic variables. 

At present, about 5–10 X-ray outbursts are detected annually for LMXBs \citep[and references therein]{ref:lmxbreview}, but only 2–3 are optically bright enough to allow for meaningful follow-up observations. However, upcoming capabilities from Rubin LSST, including its deeper magnitude limits and access to redder filters, are expected to greatly improve our ability to detect fainter or more heavily obscured optical counterparts. To date, only the brightest systems are just the visible tip of the iceberg that have been observed in optical wavelengths with great detail. The future expansion of monitoring will be key to building a complete picture of the full population and their outburst characteristics.

Optical synoptic surveys provide a potential means of discovering outbursts from both known and new LMXBs. The Vera C. Rubin Observatory's Legacy Survey of Space and Time (LSST) is the next generation of all-sky survey that has the potential to study and monitor outburst in LMXBs for a long period of time \citep{ref:ivezic}. LSST will survey the entire southern sky in 6 photometric bands, with a spatial resolution of less than 0.7 arcseconds and to faint magnitudes (g $\leq$ 25.0 mag in 30s). More importantly, Rubin's LSST will continue for 10 years, delivering an extraordinary catalog of long-baseline light curves with typical cadences of once per few hours-nights in each filter. The telescope has had its first light in 2025, with the first images detecting asteroids, comets, pulsating stars, supernova explosions and far-off galaxies. Once the 10-yr survey is underway, it is expected to collect $\sim$ 20 TB of data per night, with the users of the telescope receiving almost $\sim$ 10 million alerts. The huge amount of data as well as alerts exceeds any survey to date. This work will help in establishing new alerts as well as improving existing alerts for LMXB outburst \citep{ref:ivezic}. 

We therefore set out to investigate the prompt optical detectability of LMXB outbursts by Rubin observatory under several plausible observing strategies. Pre-outburst optical activity broadly partitions into three timescales. 1. {\it Long-term:} month- or year-long flux rise before outburst \citep{ref:dubus}. 2. {\it Fortnight:} a steeper increase in the two weeks or more before the fast rise of the outburst itself \citep{ref:ber}. This paper focuses on the third kind, 3. {\it Prompt:}, based on detection of the outburst itself due to
the fast rise and exponential decay characteristic of the light curve, the details of which are described in Section \ref{sec:xrbmetric}. In particular, we focused on exploring early detection and recovery of LMXB outburst rise light curve via forward modeling of the intrinsic outburst light curve through characteristics such as survey cadence and coverage, filter-specific magnitude depth and other observational effects.  

Our work contributes to the broader community effort to ensure the scientific return of the Rubin Observatory is maximized for as broad a range of scientific stakeholders as possible. Indeed the Rubin Observatory is somewhat unique in the degree to which it has involved the community in this effort (including, but not limited to, the development of the Rubin Simulations Framework that we utilize: see e.g. Section \ref{ssec:rubinsim}). For an overview of this community-driven process, we refer the reader to \citet[][and references therein]{bianco22}, which opens the Astrophysical Journal Focus Issue on Rubin LSST cadence and survey strategy. Several works in that issue explore recovery of transient and periodic variables at various timescales and with various spectral energy distributions (and therefore) color evolutionary profiles.  The detectability of phenomena at timescales of hours has been explored in a more general way by \citet{bellm2022}. \citet{bellm2022} introduces a simple and informative metric to evaluate how well time-domain surveys like LSST sample short timescales, finding that existing LSST cadence simulations at the time poorly sample crucial hour-to-day timescales and urges to adapt cadence strategies to address this limitation. On the other hand the ``unknown unknowns,'' phenomena whose timescales are not known before the survey commences, are discussed in \citet{li2022}. Science cases for the detection of transients and variables, are presented in the Transients \& Variable Stars (TVS) Science Collaboration Roadmap \citep{ref:hambleton}. The detectability by Rubin of {\it periodic} LMXBs via examination of optical light curves, has been explored for earlier versions of the Rubin observing strategy by \citet{ref:johnson}.

Our exploration of prompt optical detection of LMXB outbursts by Rubin up to $\sim$1 week before X-ray maximum, is complementary to these efforts. One of the aims of the investigation is to inform the LSST Survey Cadence Optimization Committee  and the community interested in LMXB science with LSST of the possible outcomes when the survey comes online. An underlying goal of our project is also to provide the community with the necessary pipeline to investigate in detail the impact of LSST observing strategies on the detection and recovery of outburst times for LMXB systems. 

The structure of the paper is as follows. Section \ref{sec:xrbmetric} describes the metrics and the developments done as part of this work. Section \ref{sec:rd} summarizes the results we obtained regarding detectability of LMXBs in outburst and discusses the implications of this work for future studies of LMXBs with Rubin. In Section \ref{sec:conclu} we conclude by providing a broad perspective of our current understanding and the possible ways to improve the XRB science outcomes from the LSST. 

\section{Methods}
\label{sec:xrbmetric}

An important challenge in LMXB outbursts is the randomness of these events. The observed sample of LMXBs represents only a small subset of the whole intrinsic Galactic population, which is expected to trace the underlying stellar mass distribution with high scatter, with contributions from both the disk and bulge components \citep{ref:grimm}. Predominantly, we discover new LMXBs when they exhibit outbursts. Furthermore, the timescale and duty cycle of these events are also random and poorly constrained (both theoretically and observationally). Thus, in this paper, we approach this problem assuming that position of LMXBs and time of their outbursts are stochastic processes, following some expected distributions, such as \citet{ref:grimm,ref:atri} for their spatial distribution and \citet{ref:chen} for outburst timescales. While this approach will not characterize the total number of outbursts and or likely LMXBs that would exhibit outbursts, it will enable us to efficiently identify optimal survey strategies for catching these outbursts early, which is vital for study of accretion in LMXBs.

\subsection{LSST operation and observation simulators}\label{ssec:rubinsim}

The LSST simulations framework \citep[\textsc{Rubin\_sim;}][]{ref:connolly}, uses a comprehensive Metric Analysis Framework (MAF) to evaluate the effectiveness of its observing strategy for diverse scientific goals. In broad overview, MAF provides a set of tools to read the \textsc{OpSim} simulated survey from a database, slice the data according to the values of single or multiple columns within the data or the spatial location of the data points, apply algorithms called metrics to each data slice and saves the results which can then be visualized and analyzed.

 A ``metric'' quantifies various aspects of observational performance, such as the number of visits per sky region, depth, cadence, and detectability of astrophysical transients. These LSST metrics are implemented using the \textsc{maf} module within \textsc{rubin\_sim}, where predefined metrics assess observation frequency and quality, while custom metrics allow for tailored scientific evaluations. For example, \texttt{CountMetric} provides the count of how many visits were in each year for the simulation , \texttt{MeanMetric} estimates the mean value of each data slice as the slicer iterates through all the data  and \texttt{RmsMetric} is used to calculate the root mean square (RMS) of data slices and is typically a measure of the stability of the observational conditions over the whole period of the simulation. Metrics operate within a ``slicing'' framework to enable filtering and selection in spatial or temporal coverage, for example with tools such as \texttt{HealpixSlicer} for slicing in sky coverage and \texttt{OneDSlicer} for filtering of time series and light curves. By leveraging these metrics, LSST ensures optimal scheduling strategies that maximize its scientific return across various areas of interest. After the metric is evaluated the resulting metric values are saved in the \texttt{MetricBundle}. The \texttt{MetricBundle} is defined by a combination of a single metric, a slicer and a constraint, which is a unique combination of operations simulation benchmark. It also saves the summary metrics to be used to generate summary statistics over those metric values as well as the resulting statistic values.


To generate observations by LSST, we use \textsc{rubin\_sim} \citep[version 2.0.0;][]{ref:connolly} and LSST Operation Simulator \textsc{Opsim} \citep[version v4.3;][]{ref:delgado}. \textsc{Opsim} generates a complete set of observational metadata for the 10 yr simulated survey of LSST. We are using the most recent version of both \textsc{rubin\_sim} and \textsc{Opsim} at the time of this writing. 

We use the \texttt{xrb\_metric}\footnote{by E. Bellm, available at \url{https://github.com/lsst/rubin_sim_notebooks/blob/main/maf/science/XRB_Metric.ipynb}} module to generate LMXB outburst light curves as they would be observed by LSST as prescribed by \textsc{rubin\_sim} and \textsc{Opsim} \citep{ref:wang}. \texttt{xrb\_metric} uses a fast rise and exponential decay (FRED) light curve template for LMXBs \citep[e.g.,][]{ref:chen}. This extended metric recovers outburst times and determines the number of detected outbursts and how many sources are observable within 0.5 days of their peak magnitude in the $ugrizy$ bands. The outbursts that are detected within this time frame of 7 days are considered to be detected in the early phase of the outbursts. By simulating XRB outbursts and adjusting for observational effects like dust extinction and distance, we assess the detectability of these outbursts under realistic survey conditions, factoring in LSST’s observing cadence and depth. The metric integrates these factors with the aim of optimizing the detection and study of XRBs in LSST. This approach models XRB outbursts using empirically derived parameters, such as orbital periods, rise and decay timescales, and peak amplitudes, within the FRED framework. We use \texttt{generate\_xrb\_pop\_slicer} to generate XRB populations, \textsc{XrbLc} class for light curve synthesis and the \textsc{XRBPopMetric} class for detectability evaluation. 


To focus on the early phase of outbursts and enable case-by-case analysis, we created a modified version of \texttt{xrb\_metric} to capture the input parameters and track individual light curves throughout the metric and Rubin simulation pipelines. This approach enables us to have a more detailed view of the ``truth'' (input parameters shaping the rise timescale and amplitude) versus what may be interpreted from the expected sparsely ``observed'' data and thus helping to determine how well we can recover outburst properties, given the current LSST survey strategy.



\subsection{Generating LSST light curves for a sample of LMXBs}

\begin{figure*}[hbt!]
    \centering
    \includegraphics[width=\linewidth]{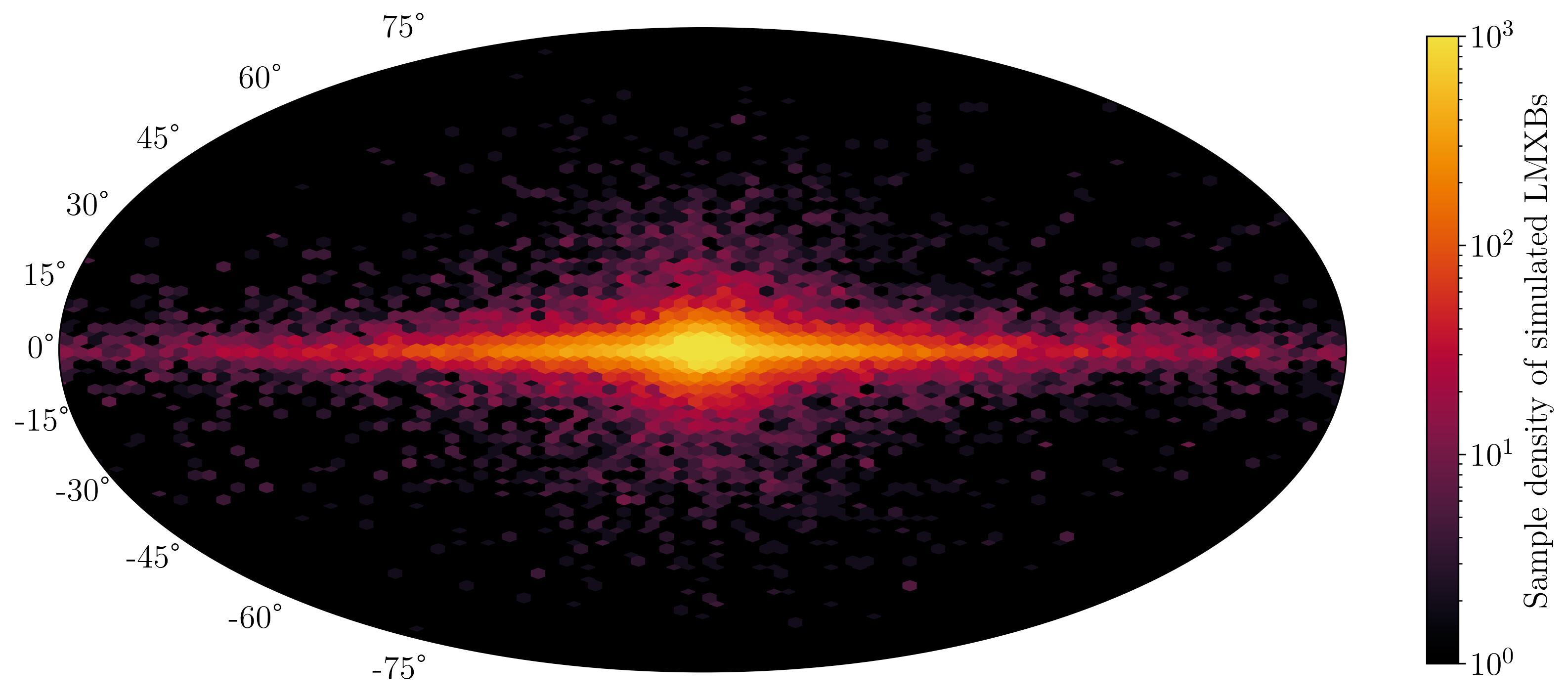}
    \includegraphics[width=\linewidth]{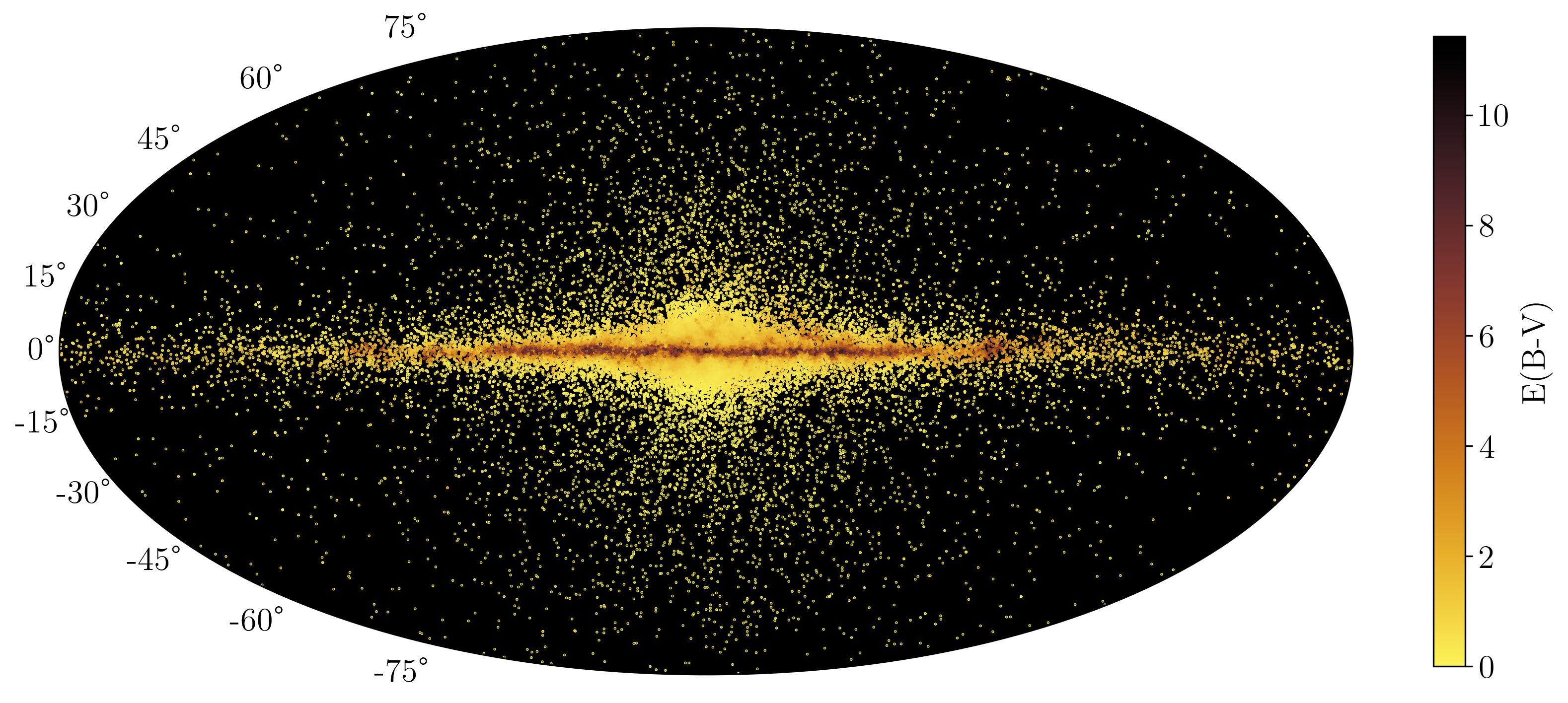}

    \caption[]{Figure highlights the intrinsic Galactic distribution of sources and the associated extinction structure. The top panel shows the sample density of simulated LMXBs used as input for the simulations, illustrating their assumed distribution relative to the Galactic plane and bulge. The bottom panel shows the corresponding line-of-sight extinction, based on the 3D extinction model and extinction map.}
    \label{fig:simgalpop}
\end{figure*}

We generate a sample of 10000 and 50000 ``events'' for deep and non deep drilling fields respectively following the Galactic distribution of LMXBs following \citet{ref:grimm, ref:atri}. The distribution of sources in our sample is shown in Figure \ref{fig:simgalpop}. The numbers of simulated events (10,000 for the deep drilling fields and 50,000 for the non-deep drilling fields) were chosen to provide statistically robust sampling of the relevant parameter space—accounting for spatial distribution, extinction, and luminosity—while remaining computationally feasible.

We also imposed a 3D extinction model on the generated sample using the \textsc{mwdust} combined extinction map \citep[\texttt{Combined19} map;][]{ref:bovy2015a}. The lower panel of Figure \ref{fig:simgalpop} shows the corresponding extinction for the sources. 


The Rubin pipeline then initializes the slicer, which generates an XRB population with the number of specified events. The slicer here includes details such as celestial coordinates, distances, extinction, and sampled system parameters (location of the events, model light curves, period, magnitude) for each simulated event. The population metric is then used to initialize the population from the MAF. It then defines the summary metrics for aggregating and analysing the results from the \texttt{XRBPopMetric}. It also calculates the mean fraction of light curves within the footprint and the average fraction of detected light curves over the entire sky. After this, the pipeline creates a bundle that encapsulates the metric and slicing information and ties it to a specific observing run, and prints out the necessary database columns required to perform the metric calculations. This setup prepares everything for executing and evaluating the metric over the specified population or sky regions.

To enable a comparison set of light curves, we also generated a second set of light curves for LMXBs assuming the coverage and survey strategy expected for the LSST deep drilling fields \citep{ref:ddf}, which are expected to be observed with much higher cadence throughout the survey. The main difference between the deep and non deep drilling fields is the cadence of observations and therefore the magnitude depths it reaches for the duration of the whole survey. The deep drilling fields of Rubin Observatory's LSST includes 5 fields chosen to maximize the multi-wavelength coverage with existing surveys. It uses a total of 6.5\% of the total survey time and receives on the order of 20,000 visits. The non-deep drilling fields are part of Rubin Observatory's LSST Wide Fast Deep (WFD) survey and composes the bulk of the survey's visits using about 80\% of the total survey time. All the non-deep drilling fields receive approximately the same number of visits per pointing which is around 800.


To explore early detection and assessing impact of survey strategy on characteristics of the outburst rise, we restrict our sample to only retain systems with detections in at least three epochs/observations in at least one filter. We then identify the first detection of the outburst ($t_{\rm{outburst}}$, defined as the earliest point $i$ in the light curve where the observed magnitude indicates a 3-$\sigma$ enhancement ($m_i<\bar{m}-3~{\rm RMSE}_m$) in each band, and the observed time of the peak of the outburst ($t_{\rm{peak}}$) in each band. These estimates provide the basic ingredients to assess early detection (pre-peak) and enable estimating the outburst rise timescale, which directly leads to estimates of viscosity in the accretion disk \citep{ref:goodwin2020}.

\section{Results}
\label{sec:rd}



In the following sections, we discuss the results obtained from running the simulations for the deep drilling and non-deep drilling fields. For both sets of fields, we detail the early science that can be extracted from the simulations as well as the recovery of the input outburst time. We also discuss the potential causes that may affect the recovery of outburst time.

\begin{figure*}[ht]
  \centering

  \begin{minipage}[b]{\linewidth}
    \centering
    \includegraphics[width=0.9\linewidth]{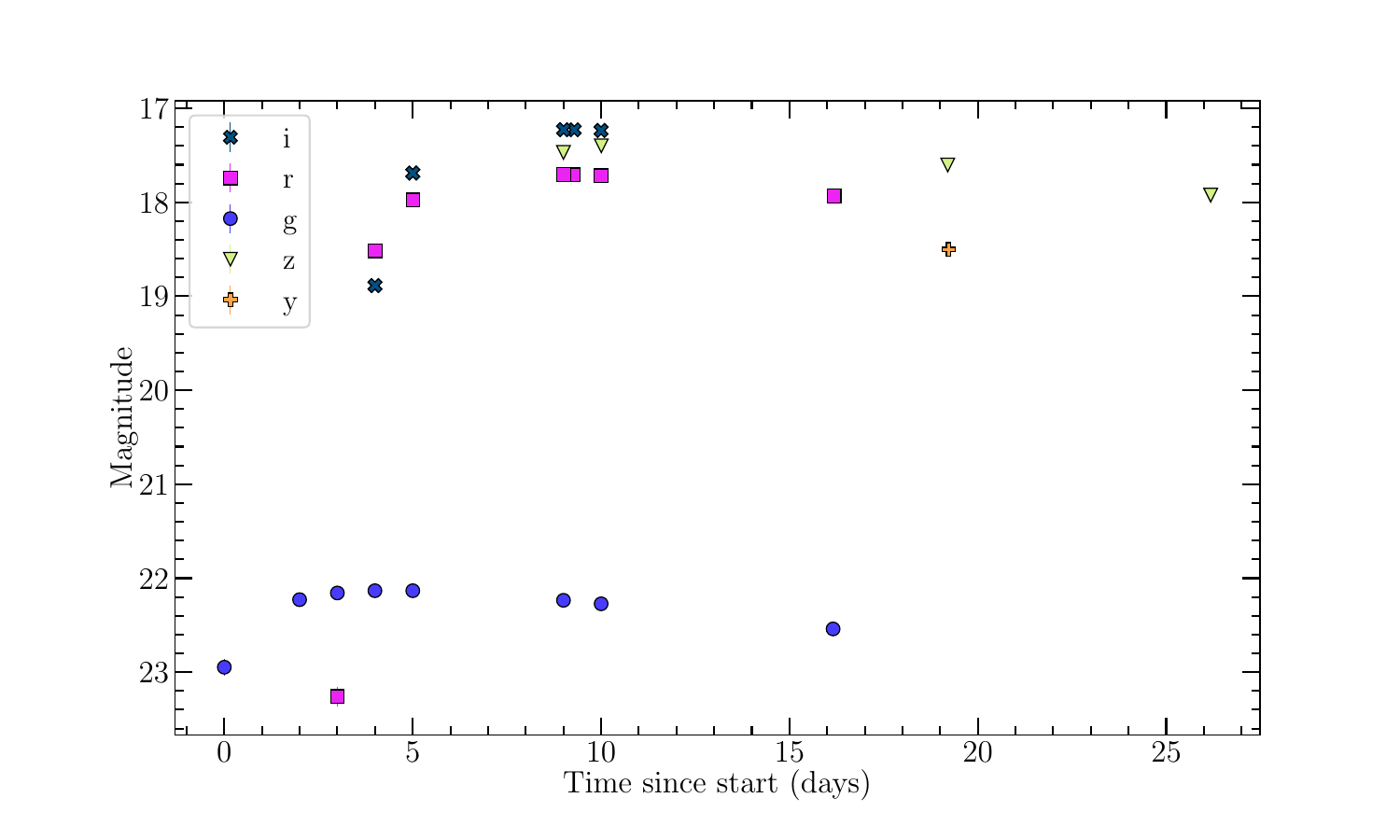} 
    \vspace{4ex}
  \end{minipage}

 \begin{minipage}[b]{\linewidth}
    \centering
    \includegraphics[width=0.95\linewidth]{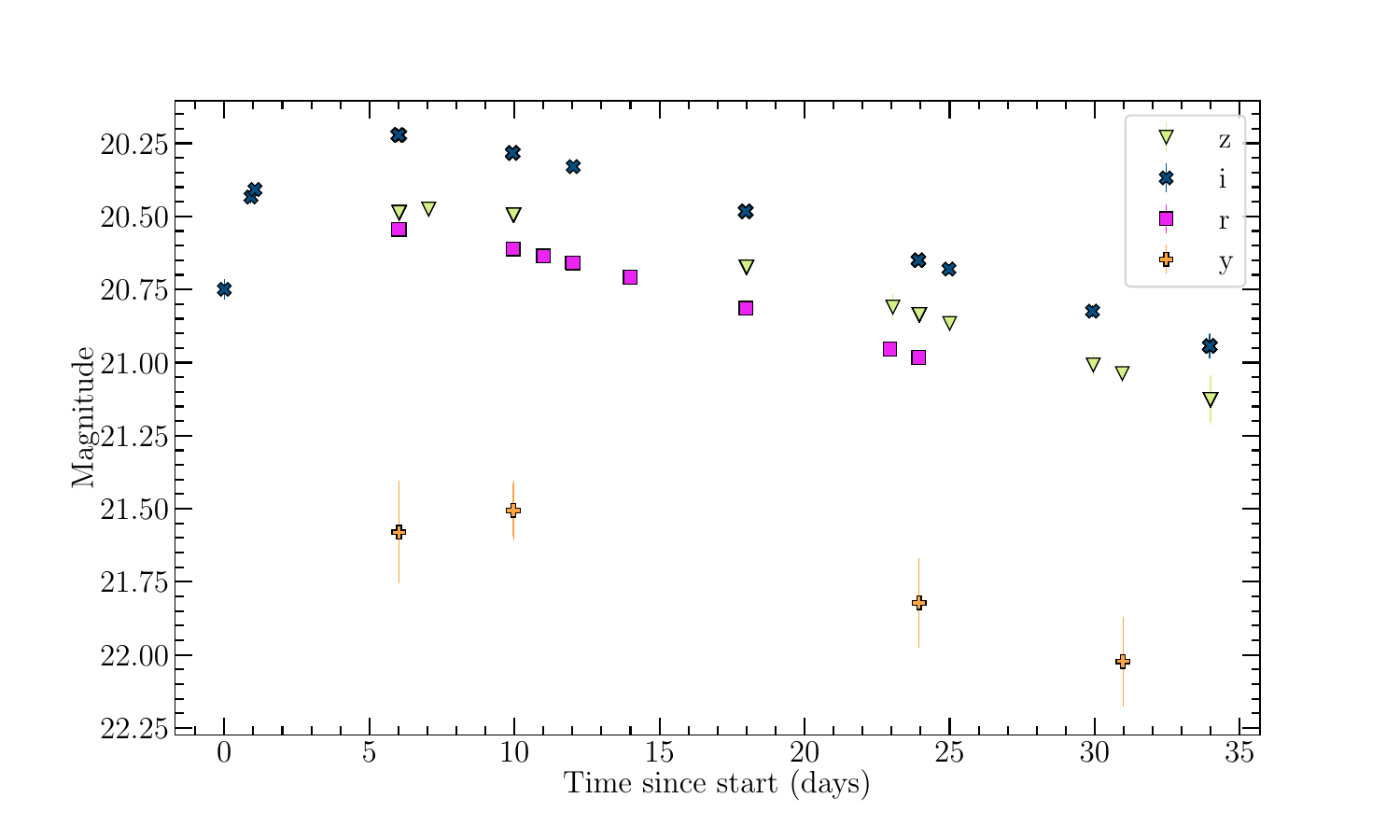} 
    \vspace{4ex}
  \end{minipage}
  
\caption{Figure shows the light curves for two sources detected in the 10 year survey. The top panel is for a source in a non deep drilling field and the bottom plot is of a source in a deep drilling field. Every coloured dot in the plot corresponds to one magnitude measurement for the light curve. The x-axis shows the time since the start of the observation in days and the y-axis shows the magnitude of the observations. The different filters shown in the figure have different coverage and extinction which affects the detectability and magnitude of the outburst.}
\label{fig:lcs}
\end{figure*}

The top and lower panel of Figure \ref{fig:lcs} shows the light curve for a source that is in a non deep and deep drilling field respectively. There are mainly three metrics that denote a detection; ``possible to detect'' which denotes the fraction of outbursts that are detected above the detection threshold, ``ever detect'' reports the fraction of events that are ever detected and ``early detect'' is the fraction of events detected twice within 7 days of the outburst start. 

Table \ref{tab:detecttable} presents the number and percentage of LMXBs expected to be detected over the entire 10-year survey for all three detection metrics for the non deep and deep drilling fields. It also shows the number of events detected when the sources are at a fixed distance of 8kpc and the extinction is varied. The approach of using a fixed distance and varying the extinction is particularly done in order to determine the role of extinction in the non-detectability of LMXB outburst. Given that the number of sources detected are not significantly more than what we detect in non-deep drilling fields, we can say that the extinction does not play a major role in the non detection of the outburst of the sources.

\begin{figure*}[ht]
\centering
 
  \begin{minipage}[b]{0.8\linewidth}
    \centering
    \includegraphics[width=\textwidth]{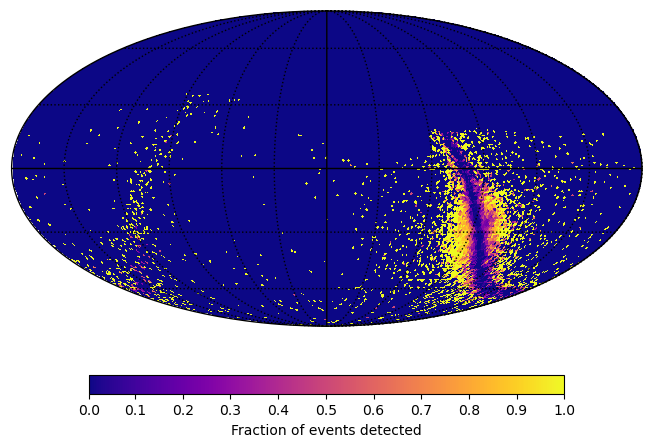} 
    \vspace{4ex}
  \end{minipage}
  
  \begin{minipage}[b]{0.8\linewidth}
    \centering
    \includegraphics[width=\linewidth]{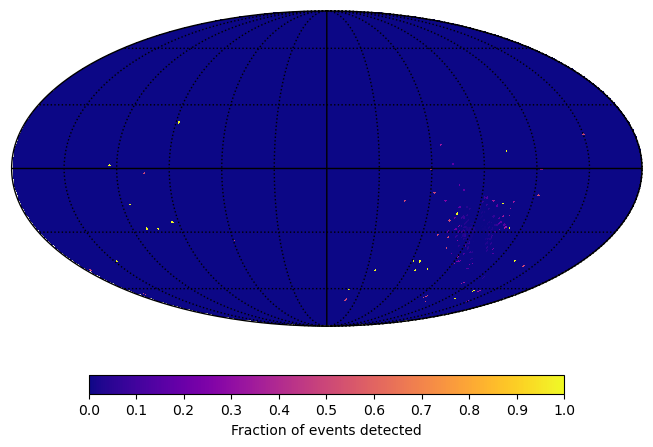} 
    \vspace{4ex}
  \end{minipage}

\caption{Sky maps illustrating the detection statistics of LMXB outbursts in the LSST baseline v4.3 10-year survey, shown in equatorial coordinates to reflect the LSST observational footprint. The top panel shows the spatial distribution of sources that are theoretically detectable above the nominal detection threshold, while the bottom panel highlights the subset of sources detected early, defined as being detected at least twice within 7 days of the outburst onset. The color bar indicates the fraction of events detected at each sky location. The reduced detectability toward the Galactic center reflects the combined effects of extinction and survey sensitivity rather than an imposed spatial exclusion.}
\label{fig:nondeepplt}
\end{figure*}

Figure \ref{fig:nondeepplt} shows the corresponding sky maps for these detections for the non deep drilling fields. 

\begin{figure*}[hbt!]
    \centering
    \includegraphics[width=\linewidth]{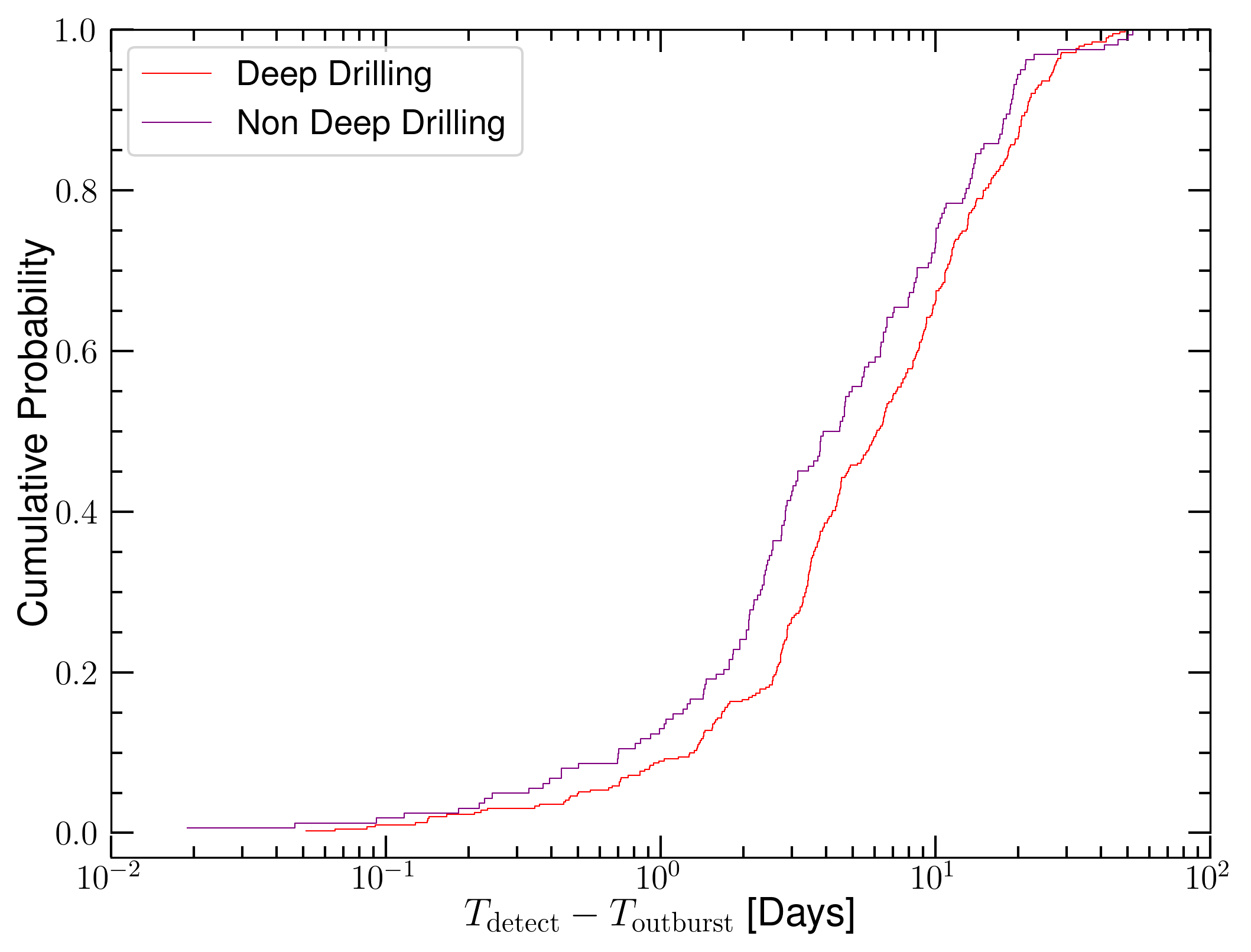}
    \caption[]{Cumulative probability distribution showing the time delay between the detection of an outburst ($\rm T_{detect}$) and its onset ($\rm T_{outburst}$) for sources observed in deep drilling fields (red curve) and non-deep drilling fields (blue curve). The plot illustrates the relative detection efficiency of the LSST survey strategy, with deep drilling fields showing a higher probability of earlier detection compared to non-deep drilling fields. This highlights the advantage of deep drilling fields in identifying outbursts more promptly, which is critical for time-sensitive astrophysical studies.}
    \label{fig:outburst}
\end{figure*}

Additionally, we examined how accurately the outburst times could be recovered compared to the true outburst times for different sources. Figure \ref{fig:outburst} illustrates the differences between the recovered and true outburst times. We also attempted to determine the time lag of outbursts between the various LSST filters. However, due to factors such as survey cadence and reduced sensitivity in certain filters, we were unable to accurately recover this lag.

\begin{table*}
\centering
\small                           
\setlength{\tabcolsep}{7pt}      
\renewcommand{\arraystretch}{1.2}

\resizebox{\textwidth}{!}{
\begin{tabular}{l*6{c}}          
\hline
\noalign{\vskip 2pt}
\textbf{}
& \multicolumn{2}{c}{\shortstack{\textbf{Deep drilling}\\\textbf{(10,000 events)}}}
& \multicolumn{2}{c}{\shortstack{\textbf{Non deep drilling}\\\textbf{(50,000 events)}}}
& \multicolumn{2}{c}{\shortstack{\textbf{Sample at 8kpc}\\\textbf{(50,000 events)}}} \\
& Total detected & \% in footprint
& Total detected & \% in footprint
& Total detected & \% in footprint \\
\hline\hline
Possible to detect & 3387 & 33 & 15399 & 30 & 16199 & 32 \\
Early detection    & 339  & 3  & 256   & 0.5 & 206   & 0.4 \\
Ever detect        & 1861 & 18 & 5163  & 10  & 4983  & 9.9 \\
\hline
\end{tabular}%
} 

\caption[]{Summary of detection statistics of sources in deep and non deep drilling fields using LSST Operation Simulator \textsc{Opsim} v4.3. The first row represents the total number and percentage of outburst that are ever above the nominal detection threshold within the 10yr survey footprint. The second row highlights the subset of outbursts that are detected twice within 7 days of the outburst start. The third row indicates the outbursts that are detected twice at any point during the survey. The columns shows the total detected outbursts and the percentage detected in the complete footprint of the survey for both deep and non deep drilling fields. The third column shows the results for a sample of sources for which the distance was kept constant at 8kpc while varying the extinction.  }
    \label{tab:detecttable}
\end{table*}

\section{Discussion}
\label{subsec:disc}
\subsection{Implications for LSST survey strategy}
From our models, deep drilling fields are able to detect a higher fraction of the light curves when compared to the non deep drilling fields for early science detections of the LMXBs. This is mainly due to more frequent temporal sampling in the deep drilling fields when compared to the non deep drilling fields. This clearly indicates that the detecting LMXBs in their outburst phase is more difficult in non deep drilling fields due to fewer number of observations for these parts of the sky. We have investigated the prospect of recovering the outburst time of LMXBs from LSST light curves. Our test case for this work is focused on the outburst time determination and recovery for LMXBs, but our results can be used more generally for assessing various proposed observatory cadence strategies, especially those relevant to the fields covered by LSST over the 10 years of the survey. 

From this work, we can see that the outburst time recovery was shown to be affected by the total number of observations in the observing strategy. The deep drilling fields resulted in more number of observations for every source in the field when compared to the number of observations in the non deep drilling fields. This resulted in the mean recovery of outburst time to be significantly better in the deep drilling field when compared to the non deep drilling field as discussed in Section \ref{sec:rd}. For both deep drilling and non deep drilling fields we were unable to recover the time lag of outbursts between the various LSST filters, which is the tracer of the accretion disk viscosity. One of the possible reasons for this could be the low number of observations for the different filters due to the survey cadence. 

As a possible recommendation to the SCOC we have tested a range of versions of Opsim for 50000 events generated as shown in Figure \ref{fig:simgalpop}. The corresponding number of detections for the baselines tested as part of this work is shown in Table \ref{tab:baselines}. We can see that the baseline one\_snap 4.0 detects more number of events when compared to the other baselines. Therefore, one of the potential ways to improve the observation frequency to get better sampling, is to plan a micro survey with the one\_snap baseline, which is similar to the v4.0 baseline but using single exposures for all visits instead of two exposure per visit for the bands. This yields a better outcome for early outburst science and can help us get the best results for this science case. The Opsim versions in the first column of the Table \ref{tab:baselines} are the changes made based on the community recommendation. More details about the observing strategy can be found at \url{https://survey-strategy.lsst.io/index.html}.

\begin{table}[]
\begin{tabular}{cccc}
\hline
Opsim versions      & Early (\%) & Ever (\%) & Possible (\%) \\ \hline
2.2           & 0.40   & 8.06 & 28.75    \\ \hline
3.4           & 0.41   & 10.39 & 28.71    \\ \hline
3.6           & 0.49   & 10.26 & 28.7    \\ \hline
4.0           & 0.53   & 10.37 & 28.71    \\ \hline
4.2           & 0.43  & 9.52 & 28.73    \\ \hline
4.3           & 0.46   & 9.64 & 28.72    \\ \hline
four\_cycle 4.0 & 0.50   & 10.34 & 28.75    \\ \hline
one\_snap 4.0   & 0.55   & 10.76 & 28.76    \\ \hline
roman         & 0.45  & 10.6 & 28.72    \\ \hline
\end{tabular}
\caption{Summary of 50000 events for different baselines for LSST. The early, ever and possible columns refer to the detection of the different detections of the events when compared to the time of the outburst. Early detection are the sources that are detected twice within the 7 days of the start of the outburst. Ever detection are the sources that are detected twice over the 10yr survey. Possible detection are the sources that are detected above the nominal detection threshold for the 10yr survey footprint.  }
\label{tab:baselines}
\end{table}

Another factor that may affect the recovery of outburst is the sensitivity of the filters. In this work, we only considered the light curves for LMXBs for which the measured magnitude of the outburst was more than the single epoch depths for every filter. This means that there may be outbursts for which observations are performed at the outburst time, but are not considered as detected as its magnitude may not be brighter than the magnitude limits for LSST filters. This is particularly seen in filter $u$, for which the single epoch magnitude depth is higher than the other filters in the most recent version of OpSim. Therefore, a combination of the sensitivity limits of LSST filters and the currently proposed survey cadence are two major factors that may be affecting the 100\% recovery of the outburst time and is majorly affecting the prospects of recovering the outburst lag between the different filters. 

One should also note that the predictions made by using the dust maps are only estimates, as the maps used represent the integrated reddening along each line of sight, therefore information on the radial change of extinction in the Galaxy is lost. Another limitation to these dust maps is their angular resolution of 6.1 arcmin. One should also note that the reddening used per field was used assuming a single pointing, corresponding to the center of the field, whereas there are potentially many different reddening values per field. 

An important factor contributing to the confirmation of the systems detected in such survey is a corresponding X-ray detection. Given that most of the systems detected would be faint, previously unknown sources, if there is no optical-X-ray correlation, it would mean that these unclassified sources potentially belong to other classes of sources such as cataclysmic variables (CVs) which show similar FRED like behavior. Bright sources detected in LSST are going to be easily detectable in large all-sky X-ray surveys. However, for the faint ones we would need a dedicated pointed follow up for such events. Given the large number of candidates it would be hard to perform a dedicated follow up of all optically detected candidates, however, these are still interesting sources that could be followed up in the long term future.

\subsection{Implication for the Galactic population of LMXBs}\label{sec:lmxbpop}
The fraction of detectable outbursts we derive in this paper enables us to put constraints on the number of LMXBs in the Galaxy as the survey progresses. However, we caution that this number is highly uncertain, as we do not fully understand the formation of these systems. For example, it is thought that dynamical formation and escape from globular clusters is thought to contribute significantly the formation of LMXBs \citep[e.g.,][]{ref:rodriguezBH, ref:gandhi20}. Observational factors such as telescope sensitivity and Galactic extinction also contribute to our incomplete understanding of LMXB numbers. Nevertheless, with these caveats, we can still build an order-of-magnitude model to estimate the number of LMXBs based on the number of outbursts detected in the LSST. Naively one can assume that the total number of detected outburst over the survey run ($N_{\rm{outburst}}$) is a Poisson process:

\begin{equation}
    N_{\rm{outburst}} \sim {\rm{Poisson}}(\lambda= f_{\rm{det}} f_{\rm{id}} \frac{T}{\tau} N_{\rm{LMXB}})
\end{equation}

Where $f_{\rm{det}}$ represents the fraction of detectable outbursts in the survey (as estimated in this paper), $f_{\rm{id}}$ represents the fraction of detected outbursts that will be identified as LMXB outbursts (e.g., assisted by brokers)\footnote{in this simple model, we assume false positive identifications are rare and negligible}, $T$ represents the length of the survey, $\tau$ represents the reoccurrence time between outbursts for an LMXB, and $N_{\rm{LMXB}}$ represents the total number of outbursting LMXBs in the Galaxy. Among these variables, $f_{\rm{det}}$ is constrained (Table~\ref{tab:detecttable}), $f_{\rm{id}}$ is currently poorly characterized, but as the LSST survey progresses, our understanding of it will be improved. $T$ is known (the length of the survey being considered). $\tau$ is highly uncertain; for example, NS-LMXBs have shown reoccurrence times between months to decades \citep{ref:heinke}, and that only includes currently known outbursting LMXBs over the age of X-ray astronomy, which is comparable to (if not shorter than) the reoccurrence time of some LMXBs. Furthermore, a single LMXB itself can show a wide range of reoccurrence times and a single value may not entirely represent the behavior.

Thus, with the caveats mentioned in mind, we can still build a simple probabilistic model, with assumptions such as $f_{\rm{det}}\sim\mathcal{U}({\rm{min}}=0.10,{\rm{max}}=0.30)$ based on our results, $f_{\rm{id}}\sim\mathcal{U}({\rm{min}}=0.10,{\rm{max}}=0.50)$ based on an ad-hoc assumption that perhaps 50\% to 90\% of LMXB outbursts are likely to be missed in classification stage (caused by factors such as poor coverage and broad range of outburst behaviors exhibited by LMXBs making classification complex), $N_{\rm{LMXB}}\sim\mathcal{U}\{{\rm{min}}=10^3,{\rm{max}}=10^5\}$ as a discrete uniform distribution bounded between twice the number of observed LMXBs \citep[e.g.,][]{ref:fortincatalog} as a minimum, and population estimates from population synthesis models \citep[e.g.,][]{ref:olejak}. The reoccurrence time of LMXBs is poorly understood. Thus, we naively adopt an ad-hoc log-normal distribution for reoccurrence time with a simplistic assumption that each LMXB only exhibits a single reoccurrence time, in two different scenarios: scenario a) that what has been observed for the reoccurrence time of NS-LMXBs is extendable to all LMXBs and is not significantly biased by LMXBs that have a long reoccurrence time. In this scenario we elicit a maximum-entropy log-normal distribution such that 95\% of all reoccurrence times are between 0.2 and 20 years, leading to $\tau/{\rm{yr}}\sim{\rm{Lognormal}}(\mu=2.0, \sigma=0.6)$; scenario b) in this scenario we assume that we have missed a somewhat larger fraction of LMXBs that have long reoccurrence time, thus we elicit a maximum-entropy log-normal distribution such that 80\% of all reoccurrence times are between 0.2 and 200 years, leading to $\tau/{\rm{yr}}\sim{\rm{Lognormal}}(\mu=4.3, \sigma=1.2)$. 

\begin{figure*}[hbt!]
    \centering
    \includegraphics[width=\linewidth]{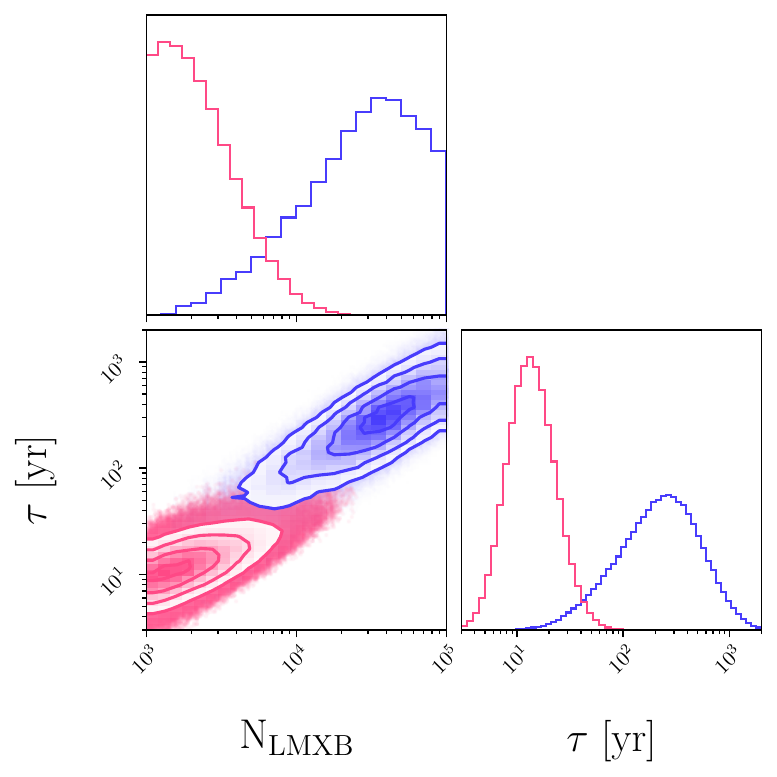}
    \caption[]{Pairwise plots demonstrating the posterior samples for the model described in \S\ref{sec:lmxbpop} in the hypothetical scenario where a total of 50 LMXB outbursts are identified over the entire life of LSST. The red contours and histograms represent scenario ``a'' (most LMXBs having shorter reoccurrence times), while the blue ones represent scenario ``b'' (a wider distribution of reoccurrence times).}
    \label{fig:lmxb_pop}
\end{figure*}

To showcase the effect of survey detection rate estimated in this work, we perform inference on $N_{\rm{LMXB}}$ with the assumptions listed above, and assuming a hypothetical survey outcome in which LSST identifies a total of 50 LMXB outbursts over its entire 10-year run. In scenario ``a", this would lead to an estimate of $\leq5,000$ in the Galaxy, while scenario ``b'' would indicate a total of 10,000-60,000 LMXBs in the Galaxy (Fig.~\ref{fig:lmxb_pop}).\footnote{Credible intervals quoted at 95\% highest-density intervals.}

\section{Conclusions}
\label{sec:conclu}

 Rubin Observatory’s LSST shows clear potential for advancing the detection and characterization of LMXBs through optical synoptic surveys. LMXB outburst are more accessible in redder filters. Compared to ZTF which has a limiting magnitude of 20.6 (see \citet{ref:ztf}) we find 60\% of our events have a peak magnitude below this detection limit (in r) demonstrating the clear potential of LSST in this domain. Furthermore, LSST's depth and multi-band coverage can potentially increase the number of optically detected candidates and add to early outburst science already being done with the dedicated alerting and follow up efforts such as X-ray Binary New Early Warning System, XB-NEWS \citep{ref:xbnews}. XB-NEWS have proven valuable in early warning of these outbursts and have already started producing useful science (see e.g., \citealt{ref:sanna,ref:saikia}). We explored the detectability of these systems across different observing conditions by simulating LMXB outbursts using a FRED model, focusing on two types of fields: deep drilling and non-deep drilling fields. 

This work indicates that deep drilling fields are much more effective for early LMXB detection, providing higher temporal coverage and, consequently, a greater likelihood of identifying outbursts soon after they occur (within 0.5 days). This suggests that increasing observation frequency in specific fields can improve the detectability and early science returns for LMXBs and other transient sources. We have tested a range of survey baselines, the results of which is shown in Table \ref{tab:baselines} and can potentially help the LSST science community to make further changes to the survey baselines to maximise science output. 

We also investigated the recovery of outburst times and the potential to measure time lags between filters. For both field types, we observed that, although the mean recovery rate of outburst times was high, recovering time lags between filters proved challenging. The inability to detect inter-filter time lags may be attributed to limitations in LSST's survey cadence and filter sensitivity, especially for the u band with its higher single-epoch magnitude threshold. These results underscore the importance of both cadence strategy and filter sensitivity in accurately capturing LMXB outbursts and other transient events.

Beyond survey cadence and sensitivity, factors such as dust extinction and distance modulus also play roles in outburst detectability. The integrated reddening values used here provide estimates, but the lack of radial extinction information and resolution limitations of dust maps add uncertainty to these predictions. Further refinement of these parameters could enhance future studies, particularly as LSST’s cadence strategies evolve with community input.

Our work provides insights for the LSST Survey Cadence Optimization Committee and contributes a robust framework for assessing LMXB detectability and recovery potential with LSST. The flexibility of the metrics developed in this study also supports broader applications for diverse time-domain science cases within the LSST community. Future research could explore optimized cadence strategies that improve detectability and data quality across a broader range of transient phenomena, ultimately maximizing the scientific impact of the Rubin Observatory’s LSST.

\begin{acknowledgement}
    The authors thank Jay Strader and Ilya Mandel for helpful discussions. This research was supported (partially or fully) by the Australian Government through the Australian Research Council's Linkage Projects funding scheme (project LE220100007).
\end{acknowledgement}



\paragraph{Data Availability Statement}
Packages used in this work are publicly available at \url{https://github.com/lsst} and \url{https://github.com/lsst/rubin_sim}


\printendnotes

\printbibliography

@ARTICLE{ref:chen,
       author = {{Chen}, Wan and {Shrader}, C.~R. and {Livio}, Mario},
        title = "{The Properties of X-Ray and Optical Light Curves of X-Ray Novae}",
      journal = {\apj},
         year = 1997,
        month = dec,
       volume = {491},
       number = {1},
        pages = {312-338},
          doi = {10.1086/304921},
archivePrefix = {arXiv},
       eprint = {astro-ph/9707138},
 primaryClass = {astro-ph},
       adsurl = {https://ui.adsabs.harvard.edu/abs/1997ApJ...491..312C}
}

@ARTICLE{ref:russel,
       author = {{Russell}, David M. and {Bramich}, Daniel M. and {Lewis}, Fraser and {AlMannaei}, Aisha and {Al Qaissieh}, Thabet and {Al Qasim}, Ahlam and {Al Yazeedi}, Aisha and {Baglio}, Maria Cristina and {Bernardini}, Federico and {Elgalad}, Nour and {Gabuya}, Aldrin and {Lasota}, Jean-Pierre and {Palado}, Alejandro and {Roche}, Paul and {Shivkumar}, Hinna and {Udrescu}, Silviu-Marian and {Zhang}, Guobao},
        title = "{Optical precursors to X‑ray binary outbursts}",
      journal = {Astronomische Nachrichten},
         year = 2019,
        month = may,
       volume = {340},
       number = {4},
        pages = {278-283},
          doi = {10.1002/asna.201913610},
archivePrefix = {arXiv},
       eprint = {1903.04519},
 primaryClass = {astro-ph.HE},
       adsurl = {https://ui.adsabs.harvard.edu/abs/2019AN....340..278R}
}

@ARTICLE{ref:hambleton,
       author = {{Hambleton}, Kelly M. and {Bianco}, Federica B. and {Street}, Rachel and others},
        title = "{Rubin Observatory LSST Transients and Variable Stars Roadmap}",
      journal = {\pasp},
         year = 2023,
        month = oct,
       volume = {135},
       number = {1052},
          eid = {105002},
        pages = {105002},
          doi = {10.1088/1538-3873/acdb9a},
archivePrefix = {arXiv},
       eprint = {2208.04499},
 primaryClass = {astro-ph.IM},
       adsurl = {https://ui.adsabs.harvard.edu/abs/2023PASP..135j5002H}
}

@ARTICLE{ref:johnson,
       author = {{Johnson}, Michael A.~C. and {Gandhi}, Poshak and {Chapman}, Adriane P. and {Moreau}, Luc and {Charles}, Philip A. and {Clarkson}, William I. and {Hill}, Adam B.},
        title = "{Prospecting for periods with LSST - low-mass X-ray binaries as a test case}",
      journal = {\mnras},
         year = 2019,
        month = mar,
       volume = {484},
       number = {1},
        pages = {19-30},
          doi = {10.1093/mnras/sty3466},
archivePrefix = {arXiv},
       eprint = {1809.09141},
 primaryClass = {astro-ph.IM},
       adsurl = {https://ui.adsabs.harvard.edu/abs/2019MNRAS.484...19J}
}

@ARTICLE{ref:grimm,
       author = {{Grimm}, Hans-Jakob and {Gilfanov}, Marat and {Sunyaev}, Rashid},
        title = "{X-ray binaries in the Milky Way and other galaxies}",
      journal = {Chinese Journal of Astronomy and Astrophysics Supplement},
         year = 2003,
        month = dec,
       volume = {3},
        pages = {257-269},
          doi = {10.1088/1009-9271/3/S1/257},
       adsurl = {https://ui.adsabs.harvard.edu/abs/2003ChJAS...3..257G}
}

@INCOLLECTION{ref:lmxbreview,
       author = {{Bahramian}, Arash and {Degenaar}, Nathalie},
        title = "{Low-Mass X-ray Binaries}",
    booktitle = {Handbook of X-ray and Gamma-ray Astrophysics},
         year = 2023,
          eid = {120},
        pages = {120},
          doi = {10.1007/978-981-16-4544-0_94-1},
       adsurl = {https://ui.adsabs.harvard.edu/abs/2023hxga.book..120B}
}

@ARTICLE{ref:ivezic,
       author = {{Ivezi{\'c}}, {\v{Z}}eljko and {Kahn}, Steven M. and others},
        title = "{LSST: From Science Drivers to Reference Design and Anticipated Data Products}",
      journal = {\apj},
         year = 2019,
        month = mar,
       volume = {873},
       number = {2},
          eid = {111},
        pages = {111},
          doi = {10.3847/1538-4357/ab042c},
archivePrefix = {arXiv},
       eprint = {0805.2366},
 primaryClass = {astro-ph},
       adsurl = {https://ui.adsabs.harvard.edu/abs/2019ApJ...873..111I}
}

@INPROCEEDINGS{ref:connolly,
       author = {{Connolly}, Andrew J. and {Angeli}, George Z. and {Chandrasekharan}, Srinivasan and {Claver}, Charles F. and {Cook}, Kem and {Ivezic}, Zeljko and {Jones}, R. Lynne and {Krughoff}, K. Simon and {Peng}, En-Hsin and {Peterson}, John and {Petry}, Catherine and {Rasmussen}, Andrew P. and {Ridgway}, Stephen T. and {Saha}, Abhijit and {Sembroski}, Glenn and {vanderPlas}, Jacob and {Yoachim}, Peter},
        title = "{An end-to-end simulation framework for the Large Synoptic Survey Telescope}",
    booktitle = {Modeling, Systems Engineering, and Project Management for Astronomy VI},
         year = 2014,
       editor = {{Angeli}, George Z. and {Dierickx}, Philippe},
       series = {Society of Photo-Optical Instrumentation Engineers (SPIE) Conference Series},
       volume = {9150},
        month = aug,
          eid = {915014},
        pages = {915014},
          doi = {10.1117/12.2054953},
       adsurl = {https://ui.adsabs.harvard.edu/abs/2014SPIE.9150E..14C}
}

@INPROCEEDINGS{ref:delgado,
       author = {{Delgado}, Francisco and {Saha}, Abhijit and {Chandrasekharan}, Srinivasan and {Cook}, Kem and {Petry}, Catherine and {Ridgway}, Stephen},
        title = "{The LSST operations simulator}",
    booktitle = {Modeling, Systems Engineering, and Project Management for Astronomy VI},
         year = 2014,
       editor = {{Angeli}, George Z. and {Dierickx}, Philippe},
       series = {Society of Photo-Optical Instrumentation Engineers (SPIE) Conference Series},
       volume = {9150},
        month = aug,
          eid = {915015},
        pages = {915015},
          doi = {10.1117/12.2056898},
       adsurl = {https://ui.adsabs.harvard.edu/abs/2014SPIE.9150E..15D}
}

@ARTICLE{ref:kawamuro,
       author = {{Kawamuro}, T. and {Negoro}, H. and {Yoneyama}, T. and {Ueno}, S. and {Tomida}, H. and {Ishikawa}, M. and {Sugawara}, Y. and {Isobe}, N. and {Shimomukai}, R. and {Mihara}, T. and {Sugizaki}, M. and {Nakahira}, S. and {Iwakiri}, W. and {Yatabe}, F. and {Takao}, Y. and {Matsuoka}, M. and {Kawai}, N. and {Sugita}, S. and {Yoshii}, T. and {Tachibana}, Y. and {Harita}, S. and {Morita}, K. and {Yoshida}, A. and {Sakamoto}, T. and {Serino}, M. and {Kawakubo}, Y. and {Kitaoka}, Y. and {Hashimoto}, T. and {Tsunemi}, H. and {Nakajima}, M. and {Kawase}, T. and {Sakamaki}, A. and {Maruyama}, W. and {Ueda}, Y. and {Hori}, T. and {Tanimoto}, A. and {Oda}, S. and {Morita}, T. and {Yamada}, S. and {Tsuboi}, Y. and {Nakamura}, Y. and {Sasaki}, R. and {Kawai}, H. and {Sato}, T. and {Yamauchi}, M. and {Hanyu}, C. and {Hidaka}, K. and {Yamaoka}, K. and {Shidatsu}, M.},
        title = "{MAXI/GSC detection of a probable new X-ray transient MAXI J1820+070}",
      journal = {The Astronomer's Telegram},
         year = 2018,
        month = mar,
       volume = {11399},
        pages = {1},
       adsurl = {https://ui.adsabs.harvard.edu/abs/2018ATel11399....1K}
}

@ARTICLE{ref:tucker,
       author = {{Tucker}, M.~A. and {Shappee}, B.~J. and {Holoien}, T.~W. -S. and {Auchettl}, K. and {Strader}, J. and {Stanek}, K.~Z. and {Kochanek}, C.~S. and {Bahramian}, A. and {ASAS-SN} and {Dong}, Subo and {Prieto}, J.~L. and {Shields}, J. and {Thompson}, Todd A. and {Beacom}, John F. and {Chomiuk}, L. and {ATLAS} and {Denneau}, L. and {Flewelling}, H. and {Heinze}, A.~N. and {Smith}, K.~W. and {Stalder}, B. and {Tonry}, J.~L. and {Weiland}, H. and {Rest}, A. and {Huber}, M.~E. and {Rowan}, D.~M. and {Dage}, K.},
        title = "{ASASSN-18ey: The Rise of a New Black Hole X-Ray Binary}",
      journal = {\apjl},
         year = 2018,
        month = nov,
       volume = {867},
       number = {1},
          eid = {L9},
        pages = {L9},
          doi = {10.3847/2041-8213/aae88a},
archivePrefix = {arXiv},
       eprint = {1808.07875},
 primaryClass = {astro-ph.HE},
       adsurl = {https://ui.adsabs.harvard.edu/abs/2018ApJ...867L...9T}
}

@ARTICLE{ref:heinke,
       author = {{Heinke}, Craig O. and {Zheng}, Junwen and {Maccarone}, Thomas J. and {Degenaar}, Nathalie and {Bahramian}, Arash and {Sivakoff}, Gregory R.},
        title = "{Catalog of outbursts of neutron star LMXBs}",
      journal = {arXiv e-prints},
         year = 2024,
        month = jul,
          eid = {arXiv:2407.18867},
        pages = {arXiv:2407.18867},
          doi = {10.48550/arXiv.2407.18867},
archivePrefix = {arXiv},
       eprint = {2407.18867},
 primaryClass = {astro-ph.HE},
       adsurl = {https://ui.adsabs.harvard.edu/abs/2024arXiv240718867H}
}

@ARTICLE{ref:tetarenko,
       author = {{Tetarenko}, B.~E. and {Sivakoff}, G.~R. and {Heinke}, C.~O. and {Gladstone}, J.~C.},
        title = "{WATCHDOG: A Comprehensive All-sky Database of Galactic Black Hole X-ray Binaries}",
      journal = {\apjs},
         year = 2016,
        month = feb,
       volume = {222},
       number = {2},
          eid = {15},
        pages = {15},
          doi = {10.3847/0067-0049/222/2/15},
archivePrefix = {arXiv},
       eprint = {1512.00778},
 primaryClass = {astro-ph.HE},
       adsurl = {https://ui.adsabs.harvard.edu/abs/2016ApJS..222...15T}
}

@ARTICLE{ref:goodwin2020,
       author = {{Goodwin}, A.~J. and {Russell}, D.~M. and {Galloway}, D.~K. and {Baglio}, M.~C. and {Parikh}, A.~S. and {Buckley}, D.~A.~H. and {Homan}, J. and {Bramich}, D.~M. and {in't Zand}, J.~J.~M. and {Heinke}, C.~O. and {Kotze}, E.~J. and {de Martino}, D. and {Papitto}, A. and {Lewis}, F. and {Wijnands}, R.},
        title = "{Enhanced optical activity 12 d before X-ray activity, and a 4 d X-ray delay during outburst rise, in a low-mass X-ray binary}",
      journal = {\mnras},
         year = 2020,
        month = nov,
       volume = {498},
       number = {3},
        pages = {3429-3439},
          doi = {10.1093/mnras/staa2588},
archivePrefix = {arXiv},
       eprint = {2006.02872},
 primaryClass = {astro-ph.HE},
       adsurl = {https://ui.adsabs.harvard.edu/abs/2020MNRAS.498.3429G}
}

@ARTICLE{ref:bovy2015a,
       author = {{Bovy}, Jo and {Rix}, Hans-Walter and {Green}, Gregory M. and {Schlafly}, Edward F. and {Finkbeiner}, Douglas P.},
        title = "{On Galactic Density Modeling in the Presence of Dust Extinction}",
      journal = {\apj},
         year = 2016,
        month = feb,
       volume = {818},
       number = {2},
          eid = {130},
        pages = {130},
          doi = {10.3847/0004-637X/818/2/130},
archivePrefix = {arXiv},
       eprint = {1509.06751},
 primaryClass = {astro-ph.GA},
       adsurl = {https://ui.adsabs.harvard.edu/abs/2016ApJ...818..130B}
}

@INPROCEEDINGS{ref:ddf,
       author = {{Weiner}, Frederica},
        title = "{LSST Deep Drilling Field Program}",
    booktitle = {American Astronomical Society Meeting Abstracts \#234},
         year = 2019,
       series = {American Astronomical Society Meeting Abstracts},
       volume = {234},
        month = jun,
          eid = {222.01},
        pages = {222.01},
       adsurl = {https://ui.adsabs.harvard.edu/abs/2019AAS...23422201W}
}

@ARTICLE{ref:wang,
       author = {{Wang}, Yuankun and {Bellm}, Eric C. and {Crossland}, Allison and {Clarkson}, William I. and {Mazzi}, Alessandro and {Riddle}, Reed and {Laher}, Russ R. and {Rusholme}, Ben},
        title = "{An Optical Search for New Outbursting Low Mass X-Ray Binaries}",
      journal = {\apj},
         year = 2024,
        month = feb,
       volume = {962},
       number = {1},
          eid = {91},
        pages = {91},
          doi = {10.3847/1538-4357/ad0fe4},
archivePrefix = {arXiv},
       eprint = {2311.18150},
 primaryClass = {astro-ph.HE},
       adsurl = {https://ui.adsabs.harvard.edu/abs/2024ApJ...962...91W}
}

@ARTICLE{ref:atri,
       author = {{Atri}, P. and {Miller-Jones}, J.~C.~A. and {Bahramian}, A. and {Plotkin}, R.~M. and {Jonker}, P.~G. and {Nelemans}, G. and {Maccarone}, T.~J. and {Sivakoff}, G.~R. and {Deller}, A.~T. and {Chaty}, S. and {Torres}, M.~A.~P. and {Horiuchi}, S. and {McCallum}, J. and {Natusch}, T. and {Phillips}, C.~J. and {Stevens}, J. and {Weston}, S.},
        title = "{Potential kick velocity distribution of black hole X-ray binaries and implications for natal kicks}",
      journal = {\mnras},
         year = 2019,
        month = nov,
       volume = {489},
       number = {3},
        pages = {3116-3134},
          doi = {10.1093/mnras/stz2335},
archivePrefix = {arXiv},
       eprint = {1908.07199},
 primaryClass = {astro-ph.HE},
       adsurl = {https://ui.adsabs.harvard.edu/abs/2019MNRAS.489.3116A}
}

@ARTICLE{ref:dubus,
       author = {{Dubus}, G. and {Hameury}, J. -M. and {Lasota}, J. -P.},
        title = "{The disc instability model for X-ray transients: Evidence for truncation and irradiation}",
      journal = {\aap},
         year = 2001,
        month = jul,
       volume = {373},
        pages = {251-271},
          doi = {10.1051/0004-6361:20010632},
archivePrefix = {arXiv},
       eprint = {astro-ph/0102237},
 primaryClass = {astro-ph},
       adsurl = {https://ui.adsabs.harvard.edu/abs/2001A&A...373..251D}
}

@ARTICLE{ref:ber,
       author = {{Bernardini}, F. and {Russell}, D.~M. and {Shaw}, A.~W. and {Lewis}, F. and {Charles}, P.~A. and {Koljonen}, K.~I.~I. and {Lasota}, J.~P. and {Casares}, J.},
        title = "{Events leading up to the 2015 June Outburst of V404 Cyg}",
      journal = {\apjl},
         year = 2016,
        month = feb,
       volume = {818},
       number = {1},
          eid = {L5},
        pages = {L5},
          doi = {10.3847/2041-8205/818/1/L5},
archivePrefix = {arXiv},
       eprint = {1601.04550},
 primaryClass = {astro-ph.HE},
       adsurl = {https://ui.adsabs.harvard.edu/abs/2016ApJ...818L...5B}
}

@ARTICLE{bianco22,
       author = {{Bianco}, Federica B. and {Ivezi{\'c}}, {\v{Z}}eljko and {Jones}, R. Lynne and {Graham}, Melissa L. and {Marshall}, Phil and {Saha}, Abhijit and {Strauss}, Michael A. and {Yoachim}, Peter and {Ribeiro}, Tiago and {Anguita}, Timo and {Bauer}, A.~E. and {Bauer}, Franz E. and {Bellm}, Eric C. and {Blum}, Robert D. and {Brandt}, William N. and {Brough}, Sarah and {Catelan}, M{\'a}rcio and {Clarkson}, William I. and {Connolly}, Andrew J. and {Gawiser}, Eric and {Gizis}, John E. and {Hlo{\v{z}}ek}, Ren{\'e}e and {Kaviraj}, Sugata and {Liu}, Charles T. and {Lochner}, Michelle and {Mahabal}, Ashish A. and {Mandelbaum}, Rachel and {McGehee}, Peregrine and {Neilsen}, Jr., Eric H. and {Olsen}, Knut A.~G. and {Peiris}, Hiranya V. and {Rhodes}, Jason and {Richards}, Gordon T. and {Ridgway}, Stephen and {Schwamb}, Megan E. and {Scolnic}, Dan and {Shemmer}, Ohad and {Slater}, Colin T. and {Slosar}, An{\v{z}}e and {Smartt}, Stephen J. and {Strader}, Jay and {Street}, Rachel and {Trilling}, David E. and {Verma}, Aprajita and {Vivas}, A.~K. and {Wechsler}, Risa H. and {Willman}, Beth},
        title = "{Optimization of the Observing Cadence for the Rubin Observatory Legacy Survey of Space and Time: A Pioneering Process of Community-focused Experimental Design}",
      journal = {\apjs},
         year = 2022,
        month = jan,
       volume = {258},
       number = {1},
          eid = {1},
        pages = {1},
          doi = {10.3847/1538-4365/ac3e72},
archivePrefix = {arXiv},
       eprint = {2108.01683},
 primaryClass = {astro-ph.IM},
       adsurl = {https://ui.adsabs.harvard.edu/abs/2022ApJS..258....1B}
}

@ARTICLE{bellm2022,
       author = {{Bellm}, Eric C. and {Burke}, Colin J. and {Coughlin}, Michael W. and {Andreoni}, Igor and {Raiteri}, Claudia M. and {Bonito}, Rosaria},
        title = "{Give Me a Few Hours: Exploring Short Timescales in Rubin Observatory Cadence Simulations}",
      journal = {\apjs},
         year = 2022,
        month = jan,
       volume = {258},
       number = {1},
          eid = {13},
        pages = {13},
          doi = {10.3847/1538-4365/ac4602},
archivePrefix = {arXiv},
       eprint = {2110.02314},
 primaryClass = {astro-ph.IM},
       adsurl = {https://ui.adsabs.harvard.edu/abs/2022ApJS..258...13B}
}

@ARTICLE{li2022,
       author = {{Li}, Xiaolong and {Ragosta}, Fabio and {Clarkson}, William I. and {Bianco}, Federica B.},
        title = "{Preparing to Discover the Unknown with Rubin LSST: Time Domain}",
      journal = {\apjs},
         year = 2022,
        month = jan,
       volume = {258},
       number = {1},
          eid = {2},
        pages = {2},
          doi = {10.3847/1538-4365/ac3bca},
archivePrefix = {arXiv},
       eprint = {2107.10281},
 primaryClass = {astro-ph.IM},
       adsurl = {https://ui.adsabs.harvard.edu/abs/2022ApJS..258....2L}
}

@ARTICLE{ref:ztf,
       author = {{Bellm}, Eric C. and {Kulkarni}, Shrinivas R. and {Graham}, Matthew J. and others},
        title = "{The Zwicky Transient Facility: System Overview, Performance, and First Results}",
      journal = {\pasp},
         year = 2019,
        month = jan,
       volume = {131},
       number = {995},
        pages = {018002},
          doi = {10.1088/1538-3873/aaecbe},
archivePrefix = {arXiv},
       eprint = {1902.01932},
 primaryClass = {astro-ph.IM},
       adsurl = {https://ui.adsabs.harvard.edu/abs/2019PASP..131a8002B}
}

@ARTICLE{ref:xbnews,
       author = {{Russell}, David M. and {Bramich}, Daniel M. and {Lewis}, Fraser and {AlMannaei}, Aisha and {Al Qaissieh}, Thabet and {Al Qasim}, Ahlam and {Al Yazeedi}, Aisha and {Baglio}, Maria Cristina and {Bernardini}, Federico and {Elgalad}, Nour and {Gabuya}, Aldrin and {Lasota}, Jean-Pierre and {Palado}, Alejandro and {Roche}, Paul and {Shivkumar}, Hinna and {Udrescu}, Silviu-Marian and {Zhang}, Guobao},
        title = "{Optical precursors to X{\ensuremath{-}}ray binary outbursts}",
      journal = {Astronomische Nachrichten},
         year = 2019,
        month = may,
       volume = {340},
       number = {4},
        pages = {278-283},
          doi = {10.1002/asna.201913610},
archivePrefix = {arXiv},
       eprint = {1903.04519},
 primaryClass = {astro-ph.HE},
       adsurl = {https://ui.adsabs.harvard.edu/abs/2019AN....340..278R}
}

@ARTICLE{ref:sanna,
       author = {{Sanna}, A. and {Illiano}, G. and {Baglio}, M.~C. and {Russell}, D.~M. and {Borghese}, A. and {Miraval Zanon}, A. and {Marino}, A. and {Riggio}, A. and {Papitto}, A. and {Alabarta}, K. and {Di Salvo}, T. and {Anitra}, A. and {Burderi}, L. and {Lewis}, F. and {Iaria}, R. and {Buckley}, D.~A.~H.},
        title = "{Flashing fast: Characterising the 2025 outburst of MAXI J1957+032}",
      journal = {\aap},
         year = 2026,
        month = feb,
       volume = {706},
          eid = {A204},
        pages = {A204},
          doi = {10.1051/0004-6361/202557859},
archivePrefix = {arXiv},
       eprint = {2512.12998},
 primaryClass = {astro-ph.HE},
       adsurl = {https://ui.adsabs.harvard.edu/abs/2026A&A...706A.204S}
}

@ARTICLE{ref:saikia,
       author = {{Saikia}, Payaswini and {Russell}, David M. and {Bramich}, D.~M. and {Alabarta}, Kevin and {Rout}, Sandeep and {Vincentelli}, Federico and {M{\'e}ndez}, Mariano and {Altamirano}, Diego and {Garc{\'\i}a}, Federico and {Baglio}, M.~C. and {Lewis}, Fraser and {Yang}, Yi-Jung},
        title = "{Exotic optical variability in the black hole X-ray binary IGR J17091-3624}",
      journal = {arXiv e-prints},
         year = 2026,
        month = jan,
          eid = {arXiv:2601.14540},
        pages = {arXiv:2601.14540},
          doi = {10.48550/arXiv.2601.14540},
archivePrefix = {arXiv},
       eprint = {2601.14540},
 primaryClass = {astro-ph.HE},
       adsurl = {https://ui.adsabs.harvard.edu/abs/2026arXiv260114540S}
}

@ARTICLE{ref:fortincatalog,
       author = {{Fortin}, F. and {Kalsi}, A. and {Garc{\'\i}a}, F. and {Simaz-Bunzel}, A. and {Chaty}, S.},
        title = "{A catalogue of low-mass X-ray binaries in the Galaxy: From the INTEGRAL to the Gaia era}",
      journal = {\aap},
         year = 2024,
        month = apr,
       volume = {684},
          eid = {A124},
        pages = {A124},
          doi = {10.1051/0004-6361/202347908},
archivePrefix = {arXiv},
       eprint = {2401.11931},
 primaryClass = {astro-ph.HE},
       adsurl = {https://ui.adsabs.harvard.edu/abs/2024A&A...684A.124F}
}

@ARTICLE{ref:olejak,
       author = {{Olejak}, A. and {Belczynski}, K. and {Bulik}, T. and {Sobolewska}, M.},
        title = "{Synthetic catalog of black holes in the Milky Way}",
      journal = {\aap},
         year = 2020,
        month = jun,
       volume = {638},
          eid = {A94},
        pages = {A94},
          doi = {10.1051/0004-6361/201936557},
archivePrefix = {arXiv},
       eprint = {1908.08775},
 primaryClass = {astro-ph.SR},
       adsurl = {https://ui.adsabs.harvard.edu/abs/2020A&A...638A..94O}
}

@ARTICLE{ref:gandhi20,
       author = {{Gandhi}, P. and {Rao}, A. and {Charles}, P.~A. and {Belczynski}, K. and {Maccarone}, T.~J. and {Arur}, K. and {Corral-Santana}, J.~M.},
        title = "{A period-dependent spatial scatter of Galactic black hole transients}",
      journal = {\mnras},
         year = 2020,
        month = jul,
       volume = {496},
       number = {1},
        pages = {L22-L27},
          doi = {10.1093/mnrasl/slaa081},
archivePrefix = {arXiv},
       eprint = {2002.00871},
 primaryClass = {astro-ph.HE},
       adsurl = {https://ui.adsabs.harvard.edu/abs/2020MNRAS.496L..22G}
}

@ARTICLE{ref:rodriguezBH,
       author = {{Rodriguez}, Carl L. and {Chatterjee}, Sourav and {Rasio}, Frederic A.},
        title = "{Binary black hole mergers from globular clusters: Masses, merger rates, and the impact of stellar evolution}",
      journal = {\prd},
         year = 2016,
        month = apr,
       volume = {93},
       number = {8},
          eid = {084029},
        pages = {084029},
          doi = {10.1103/PhysRevD.93.084029},
archivePrefix = {arXiv},
       eprint = {1602.02444},
 primaryClass = {astro-ph.HE},
       adsurl = {https://ui.adsabs.harvard.edu/abs/2016PhRvD..93h4029R}
}


\end{document}